# Five-parameter potential box with inverse square singular boundaries

A. D. Alhaidari and T. J. Taiwo

*Saudi Center for Theoretical Physics, P. O. Box 32741, Jeddah 21438, Saudi Arabia*

**Abstract**: Using the Tridiagonal Representation Approach, we obtain solutions (energy spectrum and corresponding wavefunctions) for a new five-parameter potential box with inverse square singularity at the boundaries.

**PACS**: 03.65.Ge, 03.65.Fd, 34.80.Bm, 03.65.Ca

**Keywords**: orthogonal polynomials, tridiagonal representations, recursion relation, $1/x^2$ singular potential box, bound states.

## 1. Introduction and formulation

In this article, we study the following five-parameter one-dimensional potential box

$$V(x) = \frac{1}{1-(x/L)^2}\left\{V_0 + \frac{V_L}{(x/L)^2} + \frac{V_R}{1-(x/L)^2} + V_1\left[(x/L)^2 - \frac{1}{2}\right]\right\}, \quad (1)$$

where $0 \leq x \leq L$ and $\{V_i\}$ are real parameters such that $V_L$ and $V_R$ are greater than or equal to zero. This potential has never been studied in the published literature. It is $1/x^2$ singular at the left edge of the box ($x=0$) with a singularity strength of $V_L$. At the right edge ($x=L$), it is $1/x^2$ and $1/x$ singular with a strength of $\frac{1}{4}V_R$ and $\frac{1}{2}V_0 + \frac{1}{4}V_1$, respectively. Figure 1 is a set of plots of the potential function obtained by varying one of the parameters while keeping the others fixed. It should be obvious that the spectrum of this potential consists only of an infinite number of bound states. Now, we formulate the problem and solve it using the Tridiagonal Representation Approach (TRA) [1-6]. That is, we write the wavefunction as the bounded sum $\psi(E,x) = \sum_n f_n(E)\phi_n(x)$, where $\{\phi_n(x)\}$ is a complete set of square integrable basis functions and $\{f_n(E)\}$ are proper expansion coefficients in the energy. The basis functions $\{\phi_n(x)\}$ are chosen such that the matrix representation of the wave operator is tridiagonal and symmetric. The purpose behind the tridiagonal representation requirement will become clear shortly below. If we make the coordinate transformation $y(x) = 2(x/L)^2 - 1$, then the time-independent Schrödinger equation in the new configuration space with coordinate *y* becomes

$$(H-E)|\psi\rangle = -\frac{1}{2}\left[(y')^2\frac{d^2}{dy^2} + y''\frac{d}{dy} - 2V(y) + 2E\right]|\psi\rangle = 0, \quad (2)$$

where the prime stands for the derivative with respect to *x* and we adopted the atomic units $\hbar = m = 1$. With $y \in [-1,+1]$, we can choose the following square integrable functions as basis elements for the expansion of the wavefunction

$$\phi_n(y) = A_n(1-y)^\alpha(1+y)^\beta P_n^{(\mu,\nu)}(y), \quad (3)$$

where $P_n^{(\mu,\nu)}(y)$ is the Jacobi polynomial of degree $n = 0,1,2,..$ in $y$ and we choose the normalization constant as $A_n = \sqrt{\frac{2n+\mu+\nu+1}{2^{\mu+\nu-\frac{1}{2}}}\frac{\Gamma(n+1)\Gamma(n+\mu+\nu+1)}{\Gamma(n+\nu+1)\Gamma(n+\mu+1)}}$. The parameters $\mu$ and $\nu$ are larger than $-1$ whereas the boundary conditions and square integrability (with respect to the integration measure $dx$) dictate that $y'(1-y)^{2\alpha-1}(1+y)^{2\beta-1} = (1-y)^\mu (1+y)^\nu$ giving $2\alpha = \mu+1$ and $2\beta = \nu+\frac{1}{2}$ since $y' = (2\sqrt{2}/L)\sqrt{1+y}$. It might be worthwhile giving justification for this requirement as follows. In the Jacobi basis (3), we will eventually be using the second order differential equation of the Jacobi polynomial. Since that equation begins with $(1-y^2)\frac{d^2}{dy^2}$, we rewrite Eq. (2) as

$$(H-E)|\psi\rangle = -\frac{1}{2}\frac{y'^2}{1-y^2}\left[(1-y^2)\frac{d^2}{dy^2} + \frac{y''}{y'^2}(1-y^2)\frac{d}{dy} + 2\frac{1-y^2}{y'^2}(E-V)\right]|\psi\rangle = 0. \quad (2)'$$

Therefore, the matrix elements of the wave operator in the Jacobi basis (3) will be

$$J_{n,m} = \langle \phi_n |(H-E)|\phi_m\rangle = \frac{1}{L}\int_0^L \phi_n (H-E)\phi_m dx = -\frac{1}{2L}\int_0^L \frac{y'^2}{1-y^2}\phi_n[....]\phi_m dx \quad (4)$$

$$= -\frac{1}{2L}\int_{-1}^{+1}\frac{y'}{1-y^2}\phi_n[....]\phi_m dy = -\frac{1}{2L}A_n A_m \int_{-1}^{+1}\frac{y'}{1-y^2}(1-y)^{2\alpha}(1+y)^{2\beta}P_n^{(\mu,\nu)}(y)[....]P_m^{(\mu,\nu)}(y)dy$$

where we have used the integral transform $\frac{1}{L}\int_0^L ... dx = \frac{1}{L}\int_{-1}^{+1} ... \frac{dy}{y'}$. Consequently, as required by the orthogonality of the Jacobi polynomial we obtain the above stated constraint that $\frac{y'}{1-y^2}(1-y)^{2\alpha}(1+y)^{2\beta} = (1-y)^\mu (1+y)^\nu$. Now, from Eq. (2), we can write the wave operator $J = H - E$ as

$$J = -\frac{4/L^2}{1-y}\left[(1-y^2)\frac{d^2}{dy^2} + \frac{1}{2}(1-y)\frac{d}{dy} + \varepsilon(1-y) - \frac{4u_L}{1+y} - \frac{4u_R}{1-y} - u_1 y - 2u_0\right], \quad (5)$$

where $\varepsilon = (L^2/4)E$, $u_i = (L^2/4)V_i$ and we have used $y''/(y')^2 = \frac{1/2}{1+y}$ and wrote the potential function in the new coordinate $y$ as $V(y) = \frac{2}{1-y}\left[V_0 + \frac{2V_L}{1+y} + \frac{2V_R}{1-y} + \frac{V_1}{2}y\right]$. Consequently, the action of the wave operator on the basis elements (3) is calculated as

$$J|\phi_n\rangle = -\frac{4}{L^2}A_n(1-y)^{\frac{\mu-1}{2}-\frac{1}{2}}(1+y)^{\frac{\nu+1}{2}+\frac{1}{4}} \times$$

$$\left[\frac{1}{2}\frac{\mu^2-1-8u_R}{1-y} + \frac{1}{2}\frac{\nu^2-\frac{1}{4}-8u_L}{1+y} + \varepsilon(1-y) - u_1 y - \left(n+\frac{\mu+\nu+1}{2}\right)^2 + \frac{1}{16} - 2u_0\right]P_n^{(\mu,\nu)}(x) \quad (6)$$

where we have used the differential equation of the Jacobi polynomial that reads

$$(1-y^2)\frac{d^2}{dy^2}P_n^{(\mu,\nu)} - \left[(\mu+\nu+2)y + \mu-\nu\right]\frac{d}{dy}P_n^{(\mu,\nu)} + n(n+\mu+\nu+1)P_n^{(\mu,\nu)} = 0.$$ Thus, the matrix elements of the wave operator becomes

$$J_{m,n} = \langle \phi_m | J | \phi_n \rangle = -\frac{4}{L^2}\frac{A_m A_n}{2\sqrt{2}}\int_{-1}^{+1}(1-y)^\mu (1+y)^\nu F(y) P_m^{(\mu,\nu)}(y) P_n^{(\mu,\nu)}(y) dy, \quad (4)'$$

where $F(y)$ is the expression inside the square bracket in Eq. (6).

Now, in the TRA, the matrix representation of the wave operator (4)' is required to be tridiagonal and symmetric. The recursion relation of the Jacobi polynomial and its orthogonality [7] show that this requirement is satisfied if and only if the function $F(y)$ is linear in $y$. Therefore, to eliminate the two non-linear terms $\frac{1}{1\pm y}$ in $F(y)$ the basis parameters must be chosen such that

$$\mu^2 = 1 + 8u_R \text{ and } v^2 = \tfrac{1}{4} + 8u_L. \tag{7}$$

This implies that $u_R \geq -\tfrac{1}{8}$ and $u_L \geq -\tfrac{1}{32}$, which is automatically satisfied since $V_L$ and $V_R$ are greater than or equal to zero. The constraints (7) together with the three-term recursion relation of the Jacobi polynomials, $yP_n^{(\mu,v)}(y) = \frac{v^2-\mu^2}{(2n+\mu+v)(2n+\mu+v+2)} P_n^{(\mu,v)}(y) + \frac{2(n+\mu)(n+v)}{(2n+\mu+v)(2n+\mu+v+1)} P_{n-1}^{(\mu,v)}(y) + \frac{2(n+1)(n+\mu+v+1)}{(2n+\mu+v+1)(2n+\mu+v+2)} P_{n+1}^{(\mu,v)}(y)$, and their orthogonality property, $\frac{1}{2\sqrt{2}} A_n^2 \int_{-1}^{+1} (1-y)^\mu (1-y)^v P_n^{(\mu,v)}(y) P_m^{(\mu,v)}(y) dy = \delta_{n,m}$, give the following tridiagonal and symmetric matrix representation for the wave operator (4)'

$$\frac{L^2}{4} J_{n,m} = \left[\left(n + \tfrac{\mu+v+1}{2}\right)^2 + 2u_0 - \varepsilon - \frac{1}{16} + (\varepsilon + u_1) C_n\right] \delta_{n,m}
+ (\varepsilon + u_1)(D_{n-1}\delta_{n,m+1} + D_n\delta_{n,m-1}) \tag{8}$$

where $C_n = \frac{v^2-\mu^2}{(2n+\mu+v)(2n+\mu+v+2)}$ and $D_n = \frac{2}{2n+\mu+v+2}\sqrt{\frac{(n+1)(n+\mu+1)(n+v+1)(n+\mu+v+1)}{(2n+\mu+v+1)(2n+\mu+v+3)}}$.

Thus, the matrix wave equation $\langle \phi_n | J | \psi \rangle = \sum_m J_{n,m} f_m = 0$ gives the following symmetric three-term recursion relation for the expansion coefficients of the wave function

$$\frac{\varepsilon - 2u_0 + \tfrac{1}{16}}{\varepsilon + u_1} P_n = \left[(\varepsilon + u_1)^{-1}\left(n + \tfrac{\mu+v+1}{2}\right)^2 + C_n\right] P_n + D_{n-1}P_{n-1} + D_n P_{n+1}, \tag{9}$$

where we have written $f_n(E) = f_0(E)P_n(\varepsilon)$ making $P_0 = 1$. This relation is valid for $n = 1,2,3,...$ and it gives $P_n$ as a polynomial in $\frac{\varepsilon - 2u_0 + \tfrac{1}{16}}{\varepsilon + u_1}$ to any desired degree starting with $P_0 = 1$ and $P_1 = -D_0^{-1}\left\{C_0 + (\varepsilon + u_1)^{-1}\left[\tfrac{1}{4}(\mu+v+1)^2 - \varepsilon + 2u_0 - \tfrac{1}{16}\right]\right\}$. However, this polynomial is not found in the appropriate mathematics literature. Its analytic properties (e.g., the weight function, generating function, orthogonality, asymptotics, zeros, etc.) are yet to be derived [8]. Nonetheless, this polynomial has been encountered frequently in the physics literature while solving various problems in quantum mechanics [1-6]. Had, for example, the asymptotics of the polynomial ($\lim_{n \to \infty} P_n$) been known we could have simply read off the energy spectrum $\{\varepsilon_m\}$ from the condition that makes this asymptotics vanish at these energies [8-11]. This same condition gives the discrete version of the polynomial, $P_n(\varepsilon_m)$, that enters as expansion coefficients of the corresponding bound state wavefunction $\psi(E_m, x)$ [9]. In the absence of knowledge of the analytic properties of these polynomials, we use numerical means in the following section to obtain the energy spectrum and corresponding wavefunctions.

## 2. Energy spectrum and wavefunctions

To calculate the energy spectrum, we obtain first the Hamiltonian matrix from the wave operator matrix in (8) as $H = J|_{E=0}$. Then, the energy spectrum is calculated from the wave equation $H|\psi\rangle = E|\psi\rangle$ as the generalized eigenvalues $\{E\}$ of the matrix equation $\sum_m H_{n,m} f_m = E \sum_m \Omega_{n,m} f_m$, where

$$\Omega_{n,m} = \langle \phi_n | \phi_m \rangle = \frac{A_m A_n}{2\sqrt{2}} \int_{-1}^{+1} (1-y)^{\mu+1}(1+y)^{\nu} P_m^{(\mu,\nu)}(y) P_n^{(\mu,\nu)}(y) dy$$

$$= (1-C_n)\delta_{n,m} - D_{n-1}\delta_{n,m+1} - D_n \delta_{n,m-1}$$
(10)

Table 1 is a list of the lowest energy spectrum for a given set of values of the potential parameters and for various basis sizes. It shows the rate of convergence of these values with the size of the basis.

In Figure 2, we plot the lowest bound state wavefunctions corresponding to the physical parameters and energy spectrum of Table 1. We calculate the $m^{\text{th}}$ bound state using the sum $\psi(E_m, x) \sim \sum_{n=0}^{N-1} P_n(\varepsilon_m) \phi_n(x)$, where $N$ is some large enough integer. We note that as $N$ increases from small values, the plot becomes stable for a range of values of $N$. Then, as $N$ increases beyond some critical value, $N^c$, the sum starts to become unstable producing only oscillations that increase in number and magnitude. $N^c$ increases with the magnitude of the energy level $m$. For the physical parameters of Figure 2 and Table 1, the $N^c$ values corresponding to the bound states shown in the figure are: 14, 11, 6, 13, 16, 18, etc. Moreover, if we try to evaluate the sum at an energy other than those of the bound states then we will never reach stable results but only oscillations that increase in number and magnitude. It should also be noted that as the bound state energy level becomes high enough then the corresponding state does not "feel" the detailed structure at the bottom of the potential box. That is why the wavefunction for high energy levels look like that of the square potential well with flat bottom, $\psi(E_n, x) \sim \sin(n\pi x/L)$.

## 3. Conclusion

Using the tridiagonal representation approach, we obtained exact solutions (energy spectra and corresponding wavefunctions) for the five-parameter potential box (1), which was not studied before in the physics literature and does not belong to the conventional class of exactly solvable problems. One very important issue yet to be resolved is to derive the analytic properties of the orthogonal polynomial that satisfies the recursion relation (9). This is a pure mathematical task to be taken up, hopefully, by specialists in the field. Once such properties are obtained, all the physical features of the current system, including the energy spectrum, could be written down analytically and in closed form.

## References:

[1] A. D. Alhaidari, *An extended class of L²-series solutions of the wave equation*, Ann. Phys. **317** (2005) 152

[2]  A. D. Alhaidari, *Analytic solution of the wave equation for an electron in the field of a molecule with an electric dipole moment*, Ann. Phys. **323** (2008) 1709

[3]  A. D. Alhaidari and H. Bahlouli, *Extending the class of solvable potentials: I. The infinite potential well with a sinusoidal bottom*, J. Math. Phys. **49** (2008) 082102

[4]  A. D. Alhaidari, *Extending the class of solvable potentials: II. Screened Coulomb potential with a barrier*, Phys. Scr. **81** (2010) 025013

[5]  H. Bahlouli and A. D. Alhaidari, *Extending the class of solvable potentials: III. The hyperbolic single wave*, Phys. Scr. **81** (2010) 025008

[6]  A. D. Alhaidari, *Solution of the nonrelativistic wave equation using the tridiagonal representation approach*, J. Math. Phys. Vol. **58** (2017) 072104

[7]  R. Koekoek and R. Swarttouw, *The Askey-scheme of hypergeometric orthogonal polynomials and its q-analogues*, Reports of the Faculty of Technical Mathematics and Informatics, Number 98-17 (Delft University of Technology, Delft, 1998) pages 38-44

[8]  A. D. Alhaidari, *Orthogonal polynomials inspired by the tridiagonal representation approach*, arXiv:1703.04039 [math-ph]

[9]  A. D. Alhaidari and M. E. H. Ismail, *Quantum mechanics without potential function*, J. Math. Phys. **56** (2015) 072107

[10] K. M. Case, *Orthogonal polynomials from the viewpoint of scattering theory*, J. Math. Phys. **15** (1974) 2166

[11] J. S. Geronimo and K. M. Case, *Scattering theory and polynomials orthogonal on the real line*, Trans. Amer. Math. Soc. **258** (1980) 467494

## Table Caption

**Table 1**: The lowest part of the energy spectrum (in units of $4/L^2$) for the potential parameters $(V_0, V_1, V_L, V_R) = \left(-7, -5, \frac{1}{4}, \frac{1}{2}\right)$ in units of $4/L^2$ and for various basis sizes.

## Table 1

| $n$ | 15×15 | 20×20 | 30×30 | 100×100 |
|---|---|---|---|---|
| 0 | -12.5236133022 | -12.5236133022 | -12.5236133022 | -12.5236133022 |
| 1 | -2.2785915471 | -2.2785915471 | -2.2785915471 | -2.2785915471 |
| 2 | 5.0166151049 | 5.0166151049 | 5.0166151049 | 5.0166151049 |
| 3 | 14.7610027005 | 14.7610027005 | 14.7610027005 | 14.7610027005 |
| 4 | 27.1189258293 | 27.1189258293 | 27.1189258293 | 27.1189258293 |
| 5 | 42.0517498468 | 42.0517498468 | 42.0517498468 | 42.0517498468 |
| 6 | 59.5316472005 | 59.5316471278 | 59.5316471278 | 59.5316471278 |
| 7 | 79.5404579103 | 79.5403247627 | 79.5403247627 | 79.5403247627 |
| 8 | 102.0959989310 | 102.0652500248 | 102.0652500235 | 102.0652500235 |
| 9 | 128.2758241174 | 127.0974514206 | 127.0974494272 | 127.0974494272 |

# Figures Captions

**Fig. 1**: The potential box as a function of $x$ (with $L=2$) obtained by variation of one of the parameters while keeping the rest fixed at $(V_0,V_1,V_L,V_R)=(-4,5,2,3)$ in units of $4/L^2$: part (a) is for $V_0=\{-35,-25,-10,10\}$, part (b) is for $V_1=\{-50,-30,0,30\}$, part (c) is for $V_L=\{0,5,10,20\}$, and part (d) is for $V_R=\{0,5,10,20\}$.

**Fig. 2**: The un-normalized bound state $\psi(E_n,x)$ as a function of $x$ (with $L=2$) corresponding to the physical parameters and energy eigenvalues of Table 1.

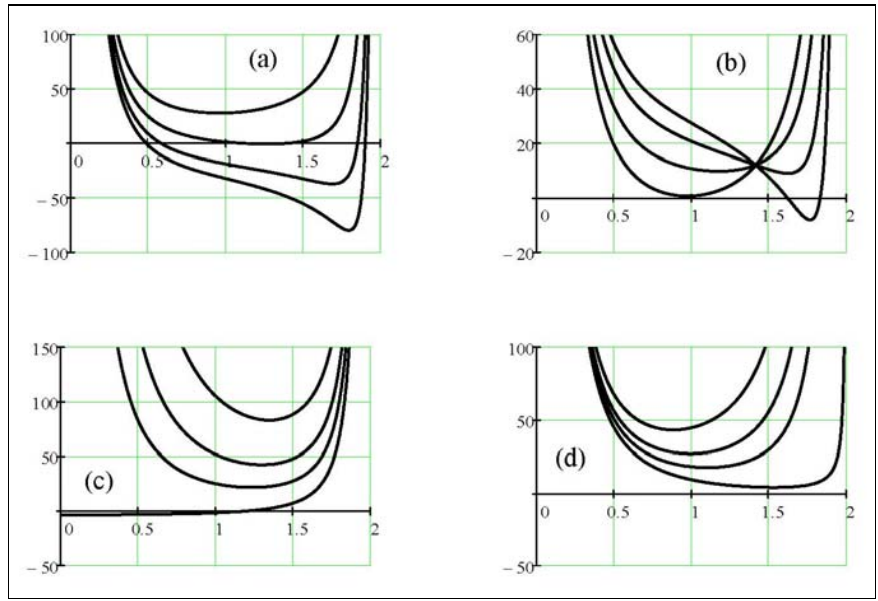

**Fig. 1**

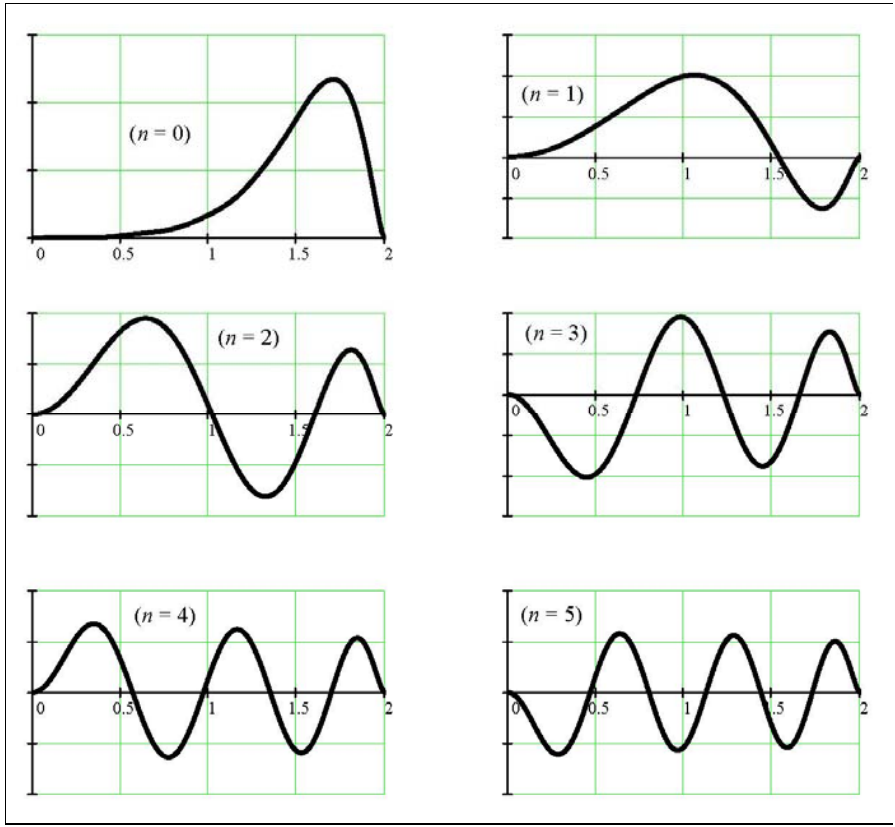

**Fig. 2**